\documentclass{article}

\usepackage{amsfonts,amssymb}
\usepackage{textcomp}
\usepackage{graphicx}
\usepackage{amsmath,amsfonts,txfonts}
\usepackage{epsfig}
\usepackage{graphicx}
\usepackage{amsmath}
\usepackage{amssymb}
\usepackage{subcaption}
\usepackage{xcolor}
\usepackage{todonotes}
\PassOptionsToPackage{numbers, compress}{natbib}

\usepackage[final]{neurips_2020}

\usepackage[utf8]{inputenc} 
\usepackage[T1]{fontenc}    
\usepackage{hyperref}       
\usepackage{url}            
\usepackage{booktabs}       
\usepackage{amsfonts}       
\usepackage{nicefrac}       
\usepackage{microtype}      
\usepackage[linesnumbered,ruled]{algorithm2e}
\usepackage{algorithmic}
\usepackage{natbib}
\usepackage{multirow}
\usepackage{diagbox}
\title{Real-time Prediction of Soft Tissue Deformations Using Data-driven Nonlinear Presurgical Simulations}

\author{
   Haolin Liu\thanks{Equal contributions. } \\
   Carnegie Mellon University \\
   Pittsburgh, PA 15213 \\
  \texttt{haolinl@andrew.cmu.edu} \\
   \And
   Ye Han\footnotemark[1] \\
   Carnegie Mellon University \\
   Pittsburgh, PA 15213 \\
  \texttt{yehan@alumni.cmu.edu} \\
   \And
   Daniel Emerson \\
   Carnegie Mellon University \\
   Pittsburgh, PA 15213 \\
  \texttt{danielem@andrew.cmu.edu} \\
   \And
   Houriyeh Majditehran \\
   Carnegie Mellon University \\
   Pittsburgh, PA 15213 \\
  \texttt{hmajdite@andrew.cmu.edu} \\
   \And
   Qi Wang \\
   Carnegie Mellon University \\
   Pittsburgh, PA 15213 \\
  \texttt{qiw2@andrew.cmu.edu} \\
   \AND
   Yoed Rabin \\
   Carnegie Mellon University \\
   Pittsburgh, PA 15213 \\
  \texttt{rabin@cmu.edu} \\
   \And
   Levent Burak Kara \\
   Carnegie Mellon University \\
   Pittsburgh, PA 15213 \\
  \texttt{lkara@cmu.edu} \\
}

\begin{document}

\maketitle

\begin{abstract} 
\label{sec:abstract}
Imaging modalities provide clinicians with real-time visualization of anatomical regions of interest (ROI) for the purpose of minimally invasive surgery. During the procedure, low-resolution image data are acquired and registered with high-resolution preoperative 3D reconstruction to guide the execution of surgical preplan. Unfortunately, due to the potential large strain and nonlinearities in the deformation of soft biological tissues, significant mismatch may be observed between ROI shapes during pre- and intra-operative imaging stages, making the surgical preplan prone to failure. In an effort to bridge the gap between the two imaging stages, this paper presents a data-driven approach based on artificial neural network for predicting the ROI deformation in real time with sparsely registered fiducial markers. For a head-and-neck tumor model with an average maximum displacement of 30 mm, the maximum surface offsets between benchmarks and predictions using the proposed approach for $98\%$ of the test cases are under $1.0$~mm, which is the typical resolution of high-quality interventional ultrasound. Each of the prediction processes takes less than $0.5$~s. With the resulting prediction accuracy and computational efficiency, the proposed approach demonstrates its potential to be clinically relevant.

\textbf{Keywords:} Soft Tissue Deformation, Real-time Deformation Prediction, Data-Driven Shape Reconstruction, Finite Element Method, Artificial Neural Network, Image-Guided Surgery
\end{abstract}

\section{Introduction}
\label{sec:intro}
In image-guided surgery~\cite{azagury2015image}, the surgical region of interest (ROI) is initially scanned under imaging modalities such as CT or MRI to create the three-dimensional (3D) reconstruction of the ROI, which is utilized afterwards for creating patient-specific surgical preplan. During the surgery, an intraoperative imaging modality such as interventional ultrasound is adopted to register the observed ROI with its pre-surgical reconstruction and to monitor the execution process of the preplan. However, due to the highly compliant nature of  biological tissues, a major challenge that often hinders the success of such image-guided surgery is the shape mismatch caused by the ROI deformation during surgery when compared with its pre-surgical reconstruction. Such a mismatch may lead to significant execution inaccuracy during the course of surgical operation and subsequently fail the preplan.

\begin{figure}[!htb]
\center{\includegraphics[width=\linewidth]
{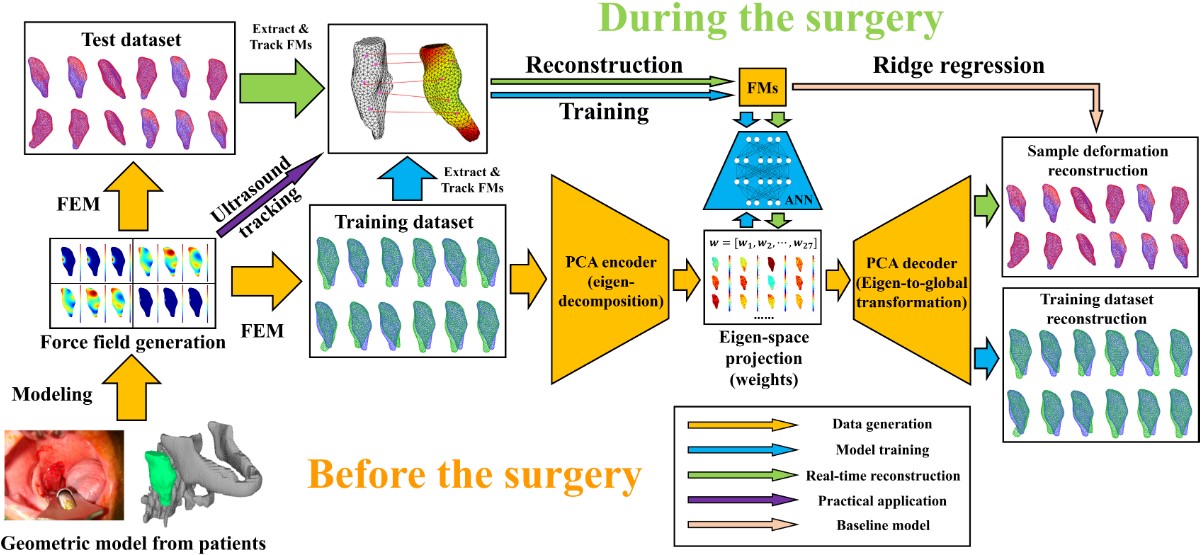}
\caption{\label{fig:pipeline} Flowchart of using proposed scheme for predicting tissue deformation during procedure. }}
\end{figure}

To address this problem, previous studies have explored different ways of accurately tracking soft tissue deformation intraoperatively. An extensively studied approach is to combine pre/intra-surgical observations with physically-based numerical simulation methods~\cite{zhang2017deformable} such as finite element method (FEM)~\cite{carter2005application,lowe2004distinctive,cash2005compensating,rucker2013mechanics,heiselman2017characterization,lunn2003nonrigid,sun2014near,khallaghi2015biomechanically,nimsky2001intraoperative}. Specifically, FEM is employed for predicting the target model's deformation by prescribing the boundary conditions (BCs) to match its intraoperative observations. By following similar deformation schemes, surgical simulation systems~\cite{monserrat2001new,zou2016new,ling2014improved} and navigation tools~\cite{jaradat2012autonomous} are developed to serve as  assistive tools to increase the guidance accuracy and to facilitate less invasive interventions. However, one limitation in the previous work is the need for repetitive forward FEM processes, which is usually computationally prohibitive especially for ROIs with complex geometric features, complicated interactions with environments, and/or nonlinear materials that require iterative solution schemes, thereby limiting their applicability under real-time scenarios. Recently, researchers have begun to combine FEM with machine learning (ML) techniques for calculating intraoperative deformation of soft biological tissues~\cite{zhou2018real,kyung2009precise,tonutti2017machine,martinez2017finite,morooka2008real,mousavi2012statistical} and accelerating the forward simulation process~\cite{yang2020simulearn}.  With the research work mentioned above, however, there still remains a question of how accurately the BCs are applied based on the observed FM tracking. In addition, it is common in practice that only limited information for spatial registration within the ROI is available, and the deformation of the tissue is often irregular and with large strains. Therefore, it is highly desirable to have a method that is able to predict the ROI deformation in real time, based on limited intraoperative registration information and maintain a desirable reconstruction performance for complex, large, nonlinear deformation scenarios. 

In this paper, a data-driven approach based on an artificial neural network (ANN) is proposed for real-time prediction of soft tissue deformations using sparsely registered landmark locations (fiducial bio-markers (FMs)) during surgery. Before surgery, the proposed approach takes input as the presurgical reconstruction of ROI geometry and a set of FM locations to create a large dataset of benchmark deformations using the eigen-decomposition of Laplace-Beltrami operator (LBO). The principle components (PCs) of the ROI deformation (displacement field) are then extracted for efficient ANN training and deformation reconstruction. After the ANN is well trained on the simulated deformation benchmarks, during the surgery, the intraoperatively registered FMs are supplied to the ANN to generate accurate deformation predictions in real time (around 0.5 s). The performance of the proposed approach is quantitatively evaluated by comparing its prediction  with the one obtained  from ridge regression. With the fast and accurate prediction results on large nonlinear deformation cases, we demonstrate our approach's potential to be clinically relevant in this presented work.

\section{Methodology}
\label{sec:method}
To illustrate our scheme, the 3D model of a head-and-neck (H\&N) squamous carcinoma is used throughout Section \ref{sec:method}. As shown in Figure \ref{fig:pipeline}, the pipeline of our proposed approach includes: (1) generating a preoperative 3D reconstruction for the patient specific surgical ROI and selecting nodes from the mesh as FMs, (2) generating a large deformation dataset using nonlinear FEM simulations and tracking the displacements of the selected FMs, (3) creating low-dimensional representations of the deformation by applying principal component analysis (PCA) to the generated dataset, (4) training an ANN on the pre-generated dataset formed by FMs' displacements and the corresponding low-dimensional representations of the deformation fields, and (5) reconstructing deformation fields of the unseen external test samples with the input of their FMs' displacements. To demonstrate the performance of our proposed ANN-based methodology, we also introduce a more classical data-driven approach based on ridge regression as the baseline method. The parameters of the H\&N tumor model as well as the nonlinearities of the FEM solver is detailed in Section \ref{sec:appendix_tumor_parameters}. 

\subsection{Dataset generation with nonlinear FEM solver}
\label{subsec:dataset_generation}
We firstly discretize the H\&N tumor geometry into tetrahedral elements and pick $n_{d}$ nodes in the tetrahedral mesh as FMs in order to track their displacements. To ensure that the FMs are sparsely and homogeneously distributed within the geometry, k-center clustering is implemented (detailed in Section \ref{sec:appendix_kcenter_clustering}). For dataset generation, we apply different force fields to the geometry's outer surface and use FEM software Abaqus as the nonlinear finite element solver to simulate the tumor's deformation in different load cases. The fixed points of the geometry are prescribed in advance, which do not overlap with the picked FMs and do not change with the force fields. The displacements of FMs $d \in \mathbb{R}^{3n_{d}\times 1}$ are tracked and recorded specifically as a one-dimensional vector for each deformed tumor sample. 

Tumors typically develop and are situated in surrounding soft tissues. To mimic this phenomenon, we constrain the synthetically generated force fields for data-set generation to be smooth and applied only to the tumor's outer surface. Therefore, we employ the LBO $\Delta_{LB}$ to the outer surface's shape function $S(x, y, z)$ to capture its curvature features and encode them in $\Delta_{LB} S$. To create smooth force fields, we firstly implement eigen-decompostition to $\Delta_{LB} S$, and then use its first 20 principal components to build LBO-reconstructed force field $f_{LBO}\in \mathbb{R}^{3n_{s}\times 1}$ with an arbitrarily generated weight vector $f_{R}$ by following the equation: $f_{LBO} = \alpha \cdot \Delta_{LB} S \cdot f_{R}$, where $n_{s}$ is the number of vertices on the outer surface, $\alpha$ is a scalar to control the magnitude of the created force fields. This approach allows arbitrary smooth force fields to be sampled and utilized to generate the training data for deformation prediction.

\subsection{Low-dimensional representation of the dataset}
\label{subsec:PCA}
After the dataset is created, the deformation of each sample can be represented as a vector $x \in \mathbb{R}^{3n_{v}\times 1}$. However, since $n_{v}$ is usually large for a mesh with good quality, it can be very difficult for a neural network to directly generate a high-dimensional output of the vector with all the nodes' deformation. To capture a lower dimensional embedding of the computed deformations fields, we apply PCA to the entire dataset to extract a linear deformation basis $P \in \mathbb{R}^{3n_{v}\times n_{w}} (n_{w}<3n_{v})$ of the dataset and effectively encode the deformation in a low-dimensional weights vector $w \in \mathbb{R}^{n_{w}\times 1}$ by following the equation $w = P^{T}x$, where $n_{w}$ is the number of principal components. By choosing a proper $n_{w}$, the information loss through PCA can be minimized and the performance of reconstruction from weight vector to the whole deformation field can be made arbitrarily accurate. For H\&N tumor model, we use 27 PCs to encode the tumor's deformation (according to the parameterization results detailed in Section \ref{sec:appendix_additional_results}). Each sample in the generated deformation dataset corresponds to a unique weight vector, which is used to train the neural network. 

\subsection{ANN training and deformation reconstruction}
\label{subsec:ANN_train}
We train the ANN with the input of FM's displacement vector $d$ and the output of the corresponding encoded weight vector $w$. We use a multilayer perceptron with 2 hidden layers as the ANN's architecture. A dataset of 2000 samples with different deformation fields is split by 8:1:1 into three subsets $G_{train}$, $G_{valid}$ and $G_{test1}$. For the H\&N tumor model, the maximum geometric dimension is rescaled to $70$ mm, and the maximum deformation of all deformed samples reaches nearly $30$ mm. Each sample in the dataset consists of $d$s and their corresponding $w$s, and the ANN is trained using the Mean Squared Error (MSE) of $w$s as the loss function. The combination of $[n_1, n_2] = [128, 64]$ (128 neurons in the first layer, 64 neurons in the second layer) is chosen as the ANN's hidden layer structure after parameterization (details of parameterization are elaborated in Section \ref{subsec:parametric_studies}). Other training process related hyperparameters are detailed in Table \ref{tab:ANN_param}. 

\begin{table}[!ht]
  \caption{Hyperparameters of benchmark dataset and ANN training for the H\&N tumor model.}
  \label{tab:ANN_param}
  \centering
  \begin{tabular}{lll}
        \toprule
        Model Parameter	& Symbol & Value \\
        \midrule
        Number of samples in $G_{train}$ & $n_{train}$ & 2000 \\
        Number of samples in $G_{test1}$ or $G_{test2}$ & $n_{test1}$ or $n_{test2}$ & 200 \\ 
        Number of neurons $[layer1, layer2]$ & $[n_{1}, n_{2}]$ & $[128, 64]$ \\
        Number of FMs & $n_{D}$ & 5 \\
        Maximum displacement in test samples & $x_{test\_max}$ & $30$ mm \\
        Training epochs & $epoch_{train}$ & 12000 \\
        Batch size & $b_{s}$ & 20 \\
        \bottomrule
  \end{tabular}
\end{table}

The trained ANN is able to predict the weight vector $w' \in \mathbb{R}^{n_{w} \times 1}$ of samples with different deformation fields with the input of corresponding FM's displacement vectors. For performance testing of the trained ANN, we establish two different test datasets $G_{test1}$ and $G_{test2}$, the latter of which is created with a completely different force field generation strategy (detailed in Section \ref{sec:appendix_force_smooth}). The deformation reconstruction of samples follows Eq. \ref{eqn:deformation_reconstruction}:

\begin{equation}
\label{eqn:deformation_reconstruction}
    \resizebox{0.9\linewidth}{!}{
    $\left[\begin{array}{c}
         | \\
         x'_{j} \\
         |
    \end{array}\right]_{n_{v}\times1} = Pw' = \left[\begin{array}{cccccc}
         | & | & \dots & | & \dots & | \\
         p_{1} & p_{2} & \dots & p_{i} & \dots & p_{n_{w}} \\
         | & | & \dots & | & \dots & |
    \end{array}\right]_{n_{v}\times n_{w}}\left[\begin{array}{c}
         w'_{1} \\
         w'_{2} \\
         \dots \\
         w'_{i} \\
         \dots \\
         w'_{n_{w}}
    \end{array}\right]_{n_{w}\times1}; j\in\{x,y,z\}$}
\end{equation}

where $x'$ denotes the reconstructed full-size deformation vector with displacements of all nodes. 

\subsection{Ridge regression and deformation reconstruction}
\label{subsec:ridge_regression}
We use ridge regression as the baseline to solve for the reconstruction weights. The function parameters are detailed in Table \ref{tab:ridgeParam} and the objective function is shown in Eq. \ref{eqn:ridgeRegression}.

\begin{table}[!ht]
    \centering
    \caption{Ridge regression parameters. }
    \begin{tabular}{lll}
        \toprule
         Parameter & Symbol & Dimension \\
         \midrule
         Nodal displacements of a single deformation benchmark & $x_{RR}$ & $\mathbb{R}^{3n_v\times1}$ \\
         Mean nodal displacement across all benchmarks & $\bar{x}_{RR}$ & $\mathbb{R}^{3n_v\times1}$ \\
         Ground truth nodal displacement of $n_{d}$ FMs & $d_{RR}$ & $\mathbb{R}^{3n_{d}\times1}$ \\
         Binary indicator matrix $\ni$ $Dx_{RR}=d_{RR}$ & $D$ & $\mathbb{R}^{3n_{d}\times3n_v}$\\
         Principal components & $P_{RR}$ & $\mathbb{R}^{3n_v\times{n_w}}$ \\
         Principal component reconstruction weights & $w_{RR}$ & $\mathbb{R}^{n_w\times1}$ \\
         \bottomrule
    \end{tabular}
    \label{tab:ridgeParam}
\end{table}

\begin{equation}
\label{eqn:ridgeRegression}
\begin{aligned}
    g(w_{RR}) &= \min_{w_{RR}} \left[a_1(||{D(P_{RR}w_{RR} + \bar{x}_{RR})-d_{RR}||}_{2}^2)+a_2w_{RR}^{T}w_{RR}\right] \\ &= \min_{w_{RR}} \left[a_1(D(P_{RR}w_{RR}+\bar{x}_{RR})-d_{RR})^T(D(P_{RR}w_{RR}+\bar{x}_{RR})-d_{RR})+a_2w_{RR}^{T}w_{RR}\right]
\end{aligned}
\end{equation}

The objective function $g(w_{RR})$ (Eq. \ref{eqn:ridgeRegression}) minimizes the difference between reconstructed and ground truth fiducial marker displacements $u_{RR} = D(P_{RR}w_{RR} + \bar{x}_{RR}) - d_{RR}$ with respect to the weight vector $w_{RR}$.  The coefficients $a_1$ and $a_2$ are weighting terms with $a_1>>a_2$. The $L_2$ regularization term $w_{RR}^{T}w_{RR}$ is weighted very lightly. The optimal ratio of ${a_2}/{a_1}={1}/{1000}$ is obtained through a parametric study.

\subsection{Evaluation Metrics}
\label{subsec:evaluation_metrics}
The evaluation metric is formulated to quantify the mismatch between the benchmarks and predicted configurations. The mean nodal offset and max nodal offset are calculated following Eq. \ref{eqn:mean_offset} and Eq. \ref{eqn:max_offset},

\begin{equation}
\label{eqn:mean_offset}
    \mathrm{Offset_{mean}} = \frac{\sum||x_{bench}^{i} - x_{pred}^{i}||_{L_{2}}}{n_{v}}
\end{equation}

\begin{equation}
\label{eqn:max_offset}
    \mathrm{Offset_{max}} = \max\limits_{i}||x_{bench}^{i} - x_{pred}^{i}||_{L_{2}}
\end{equation}

where $x_{bench}^{i}$ and $x_{pred}^{i}$ are $3\times1$ displacement vectors of vertex $i$ in benchmark and prediction and $n_{v}$ is the total number of vertices. 

\section{Results and Discussions}
\label{sec:results}
\begin{figure}[!htb]
    \centering
    \includegraphics[width=\linewidth]{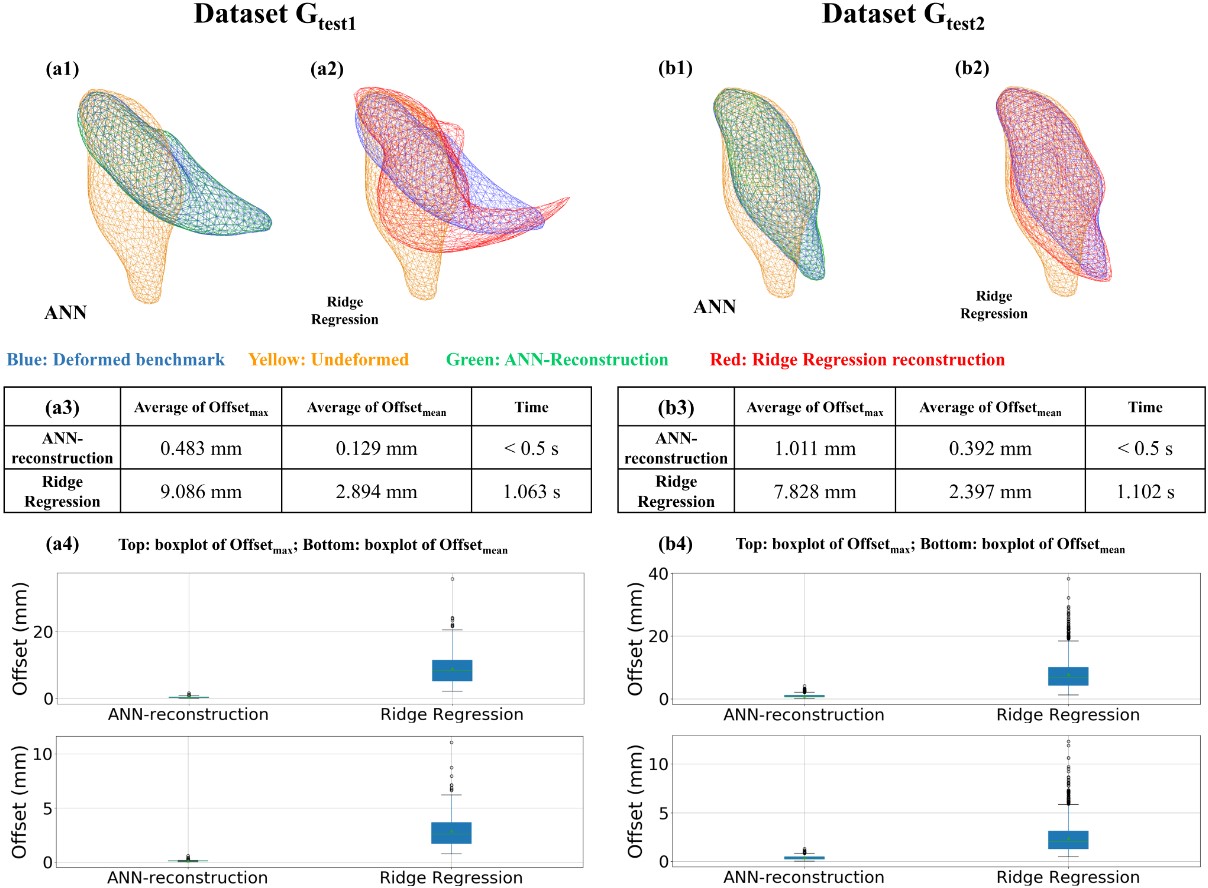}
    \caption{\label{fig:performance} Reconstruction performance of our proposed approach and the Ridge Regression method on dataset $G_{test1}$ and $G_{test2}$, respectively. (a1), (a2), (b1), (b2) visualize the matching between the undeformed (reference) configuration (in yellow), the reconstructed deformed configuration (in green or red), and the corresponding benchmark deformed configuration (in blue). (a3) and (b3) show the average of $\mathrm{Offset_{max}}$ and $\mathrm{Offset_{mean}}$ of the two approaches on $G_{test1}$ and $G_{test2}$, respectively, the boxplots of which are separately represented in (a4) and (b4). }
\end{figure}

Following the strategy introduced in Section \ref{subsec:evaluation_metrics}, we report reconstruction performances of the aforementioned methodologies for the H\&N tumor model. The ANN is implemented with PyTorch 1.6.0, and all implementations demonstrated in this paper are completed using a consumer-grade CPU (4-core Intel I7-8550U @ 1.80GHz). Parametric studies are performed on ANN training and deformation reconstruction related parameters. Overall, the proposed approach enables fast and accurate deformation reconstruction of soft tissues and outperforms the method of ridge regression, the chosen baseline model in the paper. 

\subsection{Evaluation on deformation reconstruction performance}
\label{subsec:deformation_reconstruction}
We evaluate the reconstruction performance of the proposed approach and the baseline model, respectively. For H\&N tumor, $n_{d} = 5$ FMs are chosen from the tetrahedral mesh according to the parameterization result. Figures \ref{fig:performance}(a1) and (b1) depict the comparison between the benchmark deformed configuration (in blue) and the ANN-reconstructed deformed configuration (in green) of a tested sample. Compared to the reconstruction results from the ridge regression (shown in Figures \ref{fig:performance}(a2) and (b2)), the reconstruction quality of the trained ANN is dramatically better than that of the baseline model; the boxplots of $\mathrm{Offset_{max}}$ and $\mathrm{Offset_{mean}}$ (shown in Figures \ref{fig:performance}(a4) and (b4)) also evidence this advantage of our proposed approach over the baseline method. 

The average of $\mathrm{Offset_{max}}$ and $\mathrm{Offset_{mean}}$ as well as the reconstruction time of both methods on $G_{test1}$ and $G_{test2}$ are shown in Figure \ref{fig:performance}(a3) and (b3), respectively. It can be concluded that the predicted shapes generated from our proposed approach are very close ($1.5$\% error compared to the tumor's maximum dimension) to their corresponding benchmark deformed configurations; on an unseen dataset generated by a completely different force field generation strategy of the training dataset, the trained ANN still maintains a high reconstruction quality, demonstrating the approach's robustness. In terms of the real time performance, our proposed approach generates deformed shapes within 0.5 s, which is markedly faster than the method of ridge regression and shows its capability of real-time reconstruction; the decrease of the baseline model's reconstruction speed should account for the weights searching process, which is test-dataset-dependent and sensitive to the size of the dataset. 

\subsection{Parametric Studies}
\label{subsec:parametric_studies}
For the parametric study of the ANN architecture, we explore different hidden layer structures and conduct parameterization based on their reconstruction performances on $G_{test1}$. To reduce the time cost for ANN training as well as the workload of parameterization, we only evaluate combinations of two hidden layers ($[n_1, n_2]$), each of which has a feasible neuron number list of 32, 64, 128 and 256 (resulting in a total of 16 possible combinations). Table \ref{tab:HL_parameterization} shows the results of the average of $\mathrm{Offset_{max}}$ and the elapsed time for training of all aforementioned architectures. Each of the architectures is trained three times, and the eventual result takes the average. Considering both the time efficiency of training process and the reconstruction quality, we choose the combination of [128, 64], which has relatively smallest average of $\mathrm{Offset_{max}}$ and moderate training time, as the optimal hidden layer structure. 

\begin{table}[!ht]
    \centering
    \caption{Parameterization results on ANN's two-layer hidden structures $[n_1, n_2]$. }
    \begin{tabular}{c|cccc|cccc}
        \toprule
        $[n_1, n_2]$ &
        \multicolumn{4}{c|}{Average of $\mathrm{Offset_{max}}$ (mm)} & \multicolumn{4}{c}{Training time (s)} \\
        \cmidrule{1-9}
        \diagbox{$n_1$}{$n_2$} & 32 & 64 & 128 & 256 & 32 & 64 & 128 & 256 \\
        \midrule
        32 & 0.590 & 0.538 & 0.544 & 0.574 & 747.285 & 732.081 & 788.877 & 1060.461 \\
        64 & 0.634 & 0.548 & 0.550 & 0.572 & 742.404 & 796.133 & 973.967 & 1244.056 \\
        128 & 0.893 & 0.533 & 0.539 & 0.575 & 815.583 & 1021.863 & 1255.345 & 1702.576 \\
        256 & 0.804 & 0.535 & 0.562 & 0.582 & 995.713 & 1255.417 & 1693.827 & 2668.229 \\
        \bottomrule
    \end{tabular}
    \label{tab:HL_parameterization}
\end{table}

We also conduct parametric studies on $n_{d}$ and $n_{w}$, the result of which are listed in Section \ref{sec:appendix_parameterization}. We select $n_{d} = 5$ and $n_{w} = 27$ in our implementation.  

\section{Conclusion}
\label{sec:conclusion}
In this paper, a data-driven approach for predicting intraoperative soft tissue deformation based on FM registration is developed. The proposed approach incorporates both physically-based simulations and machine learning. For an H\&N tumor model with maximum displacement of nearly $30$ mm, our approach is able to yield deformation predictions with sub-millimeter accuracy, which is the typical resolution of interventional ultrasound. Parametric studies on the number of FMs, number of PCs and ANN's hidden layer architecture are performed to characterize the proposed prediction model. Further tests on models with various geometries and topologies have demonstrated the generality of our proposed approach. After the ANN is well trained, the deformation prediction of one case takes less than $0.5$ s, showing this work’s potential for real-time applications such as intraoperative tracking of soft tissues. 

\textbf{Broader Impact. }The proposed approach can achieve real-time accurate deformation reconstruction of soft tissues. Although the pipeline is initial geometry dependent, it can still be utilized in patient-specific scenarios and assist surgeons to quickly reconstruct the deformed reconstruction by merely taking inputs from a few FMs. Overall, the presented study has demonstrated its potential to be clinically relevant.


\section*{Acknowledgements}
\label{sec:acknowledgements}
Authors would like to thank Dr. Gal Shafirstein and Roswell Park Comprehensive Cancer Center for providing H\&N tumor geometry.

\section*{Conflict of interests}
\label{sec:coi}
Authors disclaim no conflict of interests.

\bibliography{references}

\newpage
\setcounter{section}{0}
\renewcommand{\thesection}{A\arabic{section}}
\setcounter{table}{0}
\renewcommand{\thetable}{A\arabic{table}}
\setcounter{figure}{0}
\renewcommand{\thefigure}{A\arabic{figure}}
\part*{Appendix}
\label{part:appendix}
\section{Revised K-center clustering}
\label{sec:appendix_kcenter_clustering}
Section \ref{subsec:dataset_generation} introduces the revised K-center clustering algorithm for FM searching. Algorithm \ref{algrm:kcenter_clustering} shows the pseudocode detailing the implementation of the aforementioned algorithm.

\begin{algorithm}[H]
\label{algrm:kcenter_clustering}
    \SetKwInOut{Input}{Input}
    \SetKwInOut{Output}{Output}
    
    \Input{Node set $V$, initial node $v_0$, number of centers $k$, minimal distance threshold $d_{min}$}
    \Output{The list of all searched centers $V_C$}
    \Begin{
    Initialize $i \longleftarrow 0$, $V_C\longleftarrow \emptyset$, $S \longleftarrow \emptyset$; Add $v_0 \longrightarrow V_C$\;
    \While{$i < k$}{
        Set $maxAvg \longleftarrow 0$, $minVar \longleftarrow 1e5$, $v_{new} \longleftarrow \emptyset$\;
        \ForEach{$v_j\in V$}{
            \eIf{$v_j\in V_C$}{
                Skip to the next loop\;
            }{
                $S \longleftarrow \emptyset$\;
                \ForEach{$c_m\in V_C$}{
                    Compute euclidean distance between $v_j$ and $c_m$: $d_m = \mathrm{norm}(v_j, c_m)$\;
                    Add $d_m \longrightarrow S$\;
                }
                Compute average of $S$: $avg_j = \mathrm{avg}(S)$\;
                Compute minimum of $S$: $min_j = \mathrm{min}(S)$\;
                
                \eIf{$avg_j > maxAvg$ {\bf and} $min_j > d_{min}$}{
                    $v_{new} \longleftarrow v_j$\;
                    $maxAvg \longleftarrow avg_j$\;
                }{
                    Skip to the next loop\;
                }
            }
        }
        Add $v_{new} \longrightarrow V_C$; Set $v_{new} \longleftarrow \emptyset$\;
        \ForEach{$v_n\in V$}{
            \eIf{$v_n\in V_C$}{
                Skip to the next loop\;
            }{
                $S \longleftarrow \emptyset$\;
                \ForEach{$c_p\in V_C$}{
                    Compute euclidean distance between $v_n$ and $c_p$: $d_p = \mathrm{norm}(v_n, c_p)$\;
                    Add $d_p \longrightarrow S$\;
                }
                Compute variance of $S$: $var_n = \mathrm{var}(S)$\;
                Compute minimum of $S$: $min_n = \mathrm{min}(S)$\;
                
                \eIf{$var_n < minVar$ {\bf and} $min_n > d_{min}$}{
                    $v_{new} \longleftarrow v_n$\;
                    $minVar \longleftarrow var_n$\;
                }{
                    Skip to the next loop\;
                }
            }
        }
        Add $v_{new} \longrightarrow V_C$; $i \longleftarrow i + 1$\;
        }
    }
    \caption{Revised K-center clustering}
\end{algorithm}

\begin{figure}[!htb]
    \centering
    \includegraphics[width=0.9\linewidth]{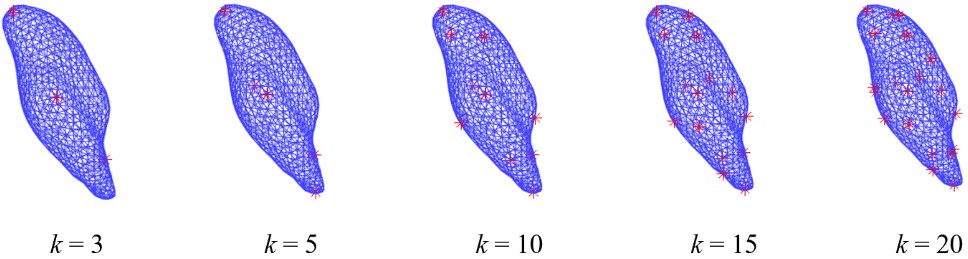}
    \caption{\label{fig:kcenters} Distribution of different numbers ($k$s) of FMs in H\&N tumor model. Starting at the initial node index of $97$, the red asterisks represent the centers obtained by the revised K-center clustering algorithm. $k = 5$ is eventually selected for implementations in this paper. }
\end{figure}

The conventional K-center clustering algorithm iteratively searches the farthest point from the pre-obtained centers within a specific closed geometry \cite{park2014greedy}. To homogeneously distribute FMs in the target geometry, the revised K-center clustering algorithm simultaneously maximizes the average and minimizes the variance of distances among the selected centers.  In this paper, the minimal distance threshold between two center points $d_{min}$ is set to 10 mm; two sub-loops are implemented in the algorithm to alternately obtain the point with the maximum distance average and the point with the minimum distance variance. With an initial point specified, the algorithm can automatically search the next best center point candidate based on the positions of pre-obtained centers and iteratively distribute the required number of center points within a closed mesh topology. Figure \ref{fig:kcenters} shows FMs captured by the algorithm with different numbers of iterations (i.e. number of points $k$) within the H\&N tumor model, evidencing that the revised K-center clustering algorithm enables the homogeneous distribution of FMs with arbitrarily specified $k$. 

\section{Introduction to modeling parameters and FEM nonlinearities}
\label{sec:appendix_tumor_parameters}
The H\&N tumor studied in the paper is CT scanned and reconstructed using the commercial Synopsys’ Simpleware™ software and tetrahedralized in Netgen~\cite{schoberl1997netgen}. The parameters of modeling and property definitions of the used H\&N model in this section are shown in Table \ref{tab:tumor_param}.

\begin{table}[!ht]
  \caption{Parameters and material properties of the H\&N tumor model. }
  \label{tab:tumor_param}
  \centering
  \begin{tabular}{lll}
        \toprule
        Model Parameter	& Symbol & H\&N Tumor \\
        \midrule
        Number of vertices & $n_{v}$ & 1158 \\
        Number of tetrahedral elements & $n_{e}$ & 8520 \\ 
        Number of vertices on the outer surface & $n_{s}$ & 760 \\
        Number of triangles on the outer surface & $n_{t}$ & 1516 \\
        Size & $dim_{x}\times dim_{y}\times dim_{z}$ & $22~mm\times 39~mm\times 70~mm$ \\
        Fixed node indices (indexed from 1) & $V_{fix}$ & [761, 1000, 1158] \\
        Fiducial marker indices (indexed from 1) & $V_{FM}$ & [97, 753, 1145, 5, 432] \\
        Young’s modulus & $E$ & $21~kPa$ \\ 
        Poisson’s ratio & $\upsilon$ & $0.45$ \\
        \bottomrule
  \end{tabular}
\end{table}

With regard to the nonlinearity of FEM solver, we include the following nonlinearities in the process of dataset generation: 

\textbf{Material nonlinearity.} We model the soft tissue as a neo-Hookean solid. Assuming the tissue is homogeneous, continuous and isotropic, the material constants can be computed based on its Young's modulus and Poisson's ratio following the strategy introduced in \cite{hu2004soft}. 

\textbf{Geometric nonlinearity.} By discretizing the simulation process, the force is divided into load increments which are gradually applied to the tumor. The transformation matrix and stiffness matrix must be recomputed at each simulation step to account for incremental changes in the geometry of the tumor. 

\textbf{Nonlinear element formulation.} Each tetrahedral element has 10 nodes (4 of them at the element's corners, 6 of them at the middle points of the element' edges), each of which possesses 6 degrees of freedom (DOFs); a quadratic displacement function is defined for each tetrahedral element. 

\section{Force field interpolation and test dataset generation}
\label{sec:appendix_force_smooth}
\begin{figure}[!htb]
    \centering
    \includegraphics[width=\linewidth]{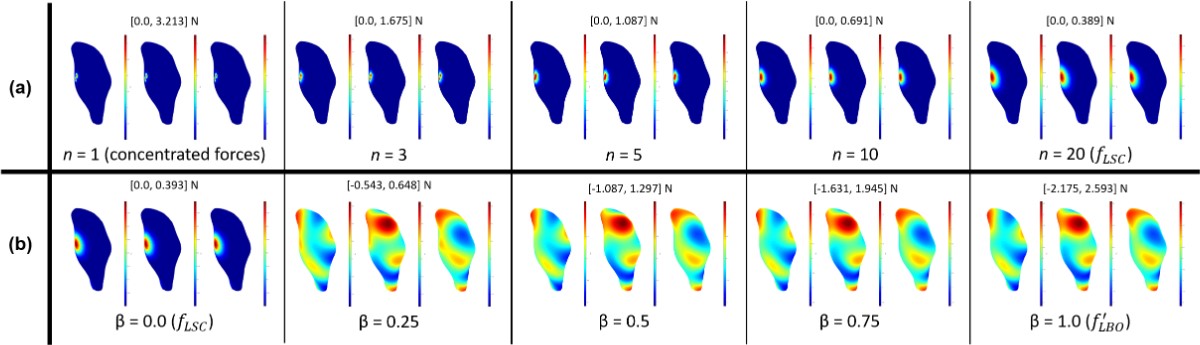}
    \caption{\label{fig:force_field_interp} Visualization of the generating process of $f_{LSC}$ and $f_{interp}$ as well as some intermediate force fields on the H\&N tumor. Row (a) depicts the process of Laplacian smoothing on three concentrated forces (in the figure, the initial magnitude of three forces: $[f_x, f_y, f_z] = [5.0, 5.0, 5.0]$ N), and $f_{LSC}$ is eventually obtained after 20 iterations of smoothing with $\gamma = 0.1$. Row (b) shows linear interpolation results $f_{interp}$, which are used to construct $G_{test2}$, with different $\beta$s between $f_{LSC}$ ($\beta = 0$) and $f'_{LBO}$ ($\beta = 1$).}
\end{figure}

The deformation reconstruction performance is firstly demonstrated with $G_{test1}$, which is a sub-dataset generated when creating $G_{train}$. To further validate that the pre-trained ANN can predict the weights vector with a high quality on samples with different boundary conditions, we establish another test dataset $G_{test2}$, which is created based on a completely different force field settings. The pipeline to generate the force fields of $G_{test2}$ is as follows: 
\begin{enumerate}
    \item Randomly pick three nodes on the outer surface and assign a random concentrated force to each node. The scalar $\alpha$, which is used to control the magnitude of $f_{LBO}$, is applied here to control the value of each concentrated force;
    \item Employ local Laplacian smoothing to each concentrated force to create a smoothed concentrated force field $f_{LSC}\in \mathbb{R}^{3n_{s}\times 1}$. In this paper, we specify the smoothing rate $\gamma = 0.1$ and iteratively smooth in a total of 20 iterations. The effect of local force smoothing is shown in Figure \ref{fig:force_field_interp}(a);
    \item Generate another LBO-reconstructed force field $f'_{LBO}\in \mathbb{R}^{3n_{s}\times 1}$ (different from $f_{LBO}$ in Section \ref{subsec:dataset_generation}) following the same strategy introduced in Section \ref{subsec:dataset_generation};
    \item Generate a new force field $f_{interp}\in \mathbb{R}^{3n_{s}\times 1}$ by linear interpolating between $f'_{LBO}$ and $f_{LSC}$ with an interpolation coefficient $\beta$. In this paper, 11 different $\beta$s are generated within the range of $[0,1]$ with an identical step size of 0.1. The results of force field interpolation is depicted in Figure \ref{fig:force_field_interp}(b).
\end{enumerate}

\section{Additional results of deformation reconstruction performance}
\label{sec:appendix_additional_results}
\begin{figure}[!htb]
    \centering
    \includegraphics[width=\linewidth]{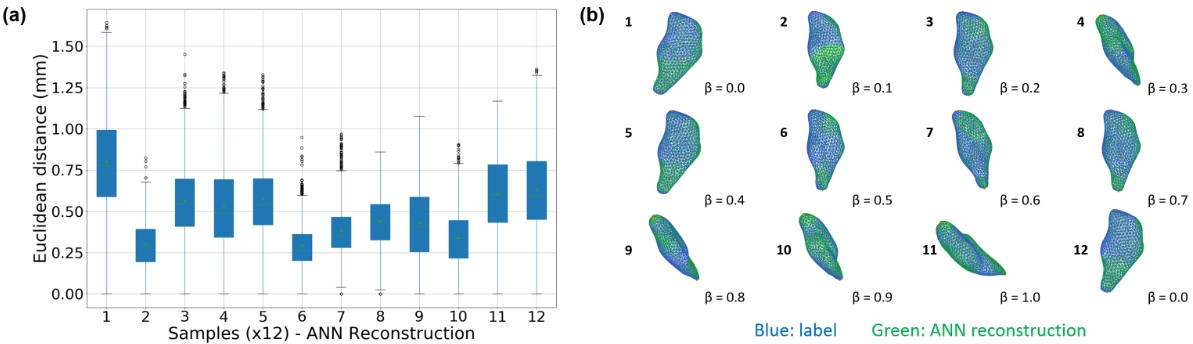}
    \caption{\label{fig:12_samples} Reconstruction results of twelve samples with different force field interpolation coefficients ($\beta$s). (a) shows the nodal offsets distribution of each sample, the average nodal mismatch of which appears to be less than $1$ mm. (b) visualizes the matching between the benchmark deformed configurations (in blue) and reconstructed configurations of our proposed approach (in green). }
\end{figure}

We provide more reconstruction results as well as some performance evaluation plots of our proposed approach here in addition to the representation in Section \ref{subsec:deformation_reconstruction}. Figure \ref{fig:12_samples} shows the result of 12 samples constructed with different $\beta$s. From Figure \ref{fig:12_samples}(a), we can clearly observe that the majority of nodal mismatch results for each sample are under $1$ mm (corresponding to an error percentage of less than $1.5$\% compared to the tumor's maximum dimension of $70$ mm), demonstrating that the reconstruction performance of trained ANN with respect to different force field constructing parameters remains extraordinarily stable and outstanding. From the visualized results in Figure \ref{fig:12_samples}(b), we can conclude that the predicted shapes generated from our proposed approach are very close to the benchmark deformed configurations. 

\begin{table}[!ht]
  \caption{Reconstruction performance of the proposed approach with respect to $\beta$. }
  \label{tab:performance_vs_beta}
  \centering
  \begin{tabular}{ccc}
        \toprule
        $\beta$	& Average of $\mathrm{Offset_{max}}$ (mm) & Average of $\mathrm{Offset_{mean}}$ (mm) \\
        \midrule
        $0$ & 1.181 & 0.452 \\
        $0.1$ & 1.085 & 0.415 \\ 
        $0.2$ & 1.016 & 0.383 \\
        $0.3$ & 0.941 & 0.370 \\
        $0.4$ & 0.932 & 0.366 \\
        $0.5$ & 0.889 & 0.350 \\ 
        $0.6$ & 0.903 & 0.356 \\
        $0.7$ & 0.971 & 0.378 \\
        $0.8$ & 0.951 & 0.379 \\
        $0.9$ & 1.089 & 0.413 \\
        $1$ & 1.167 & 0.453 \\
        \bottomrule
  \end{tabular}
\end{table}

Table \ref{tab:performance_vs_beta} summarizes results of the reconstruction performance of sample groups with respect to different $\beta$s. For each group, the result is represented in the average of $\mathrm{Offset_{mean}}$ as well as the average of $\mathrm{Offset_{max}}$ of all samples. Results further validate the point that ANN trained on LBO-reconstructed dataset can perform very well on external unseen dataset with completely different force field modes. The relative lowest reconstruction error appears at $\beta = 0.5$, the corresponding $f_{interp}$ at which has the minimal magnitude of force field and therefore the minimal magnitude of deformation. It is apparent that the smaller the deformation magnitude is, the better the reconstruction performance the ANN can achieve.

\section{Additional results of parametric studies}
\label{sec:appendix_parameterization}
\begin{figure}[!htb]
    \centering
    \includegraphics[width=\linewidth]{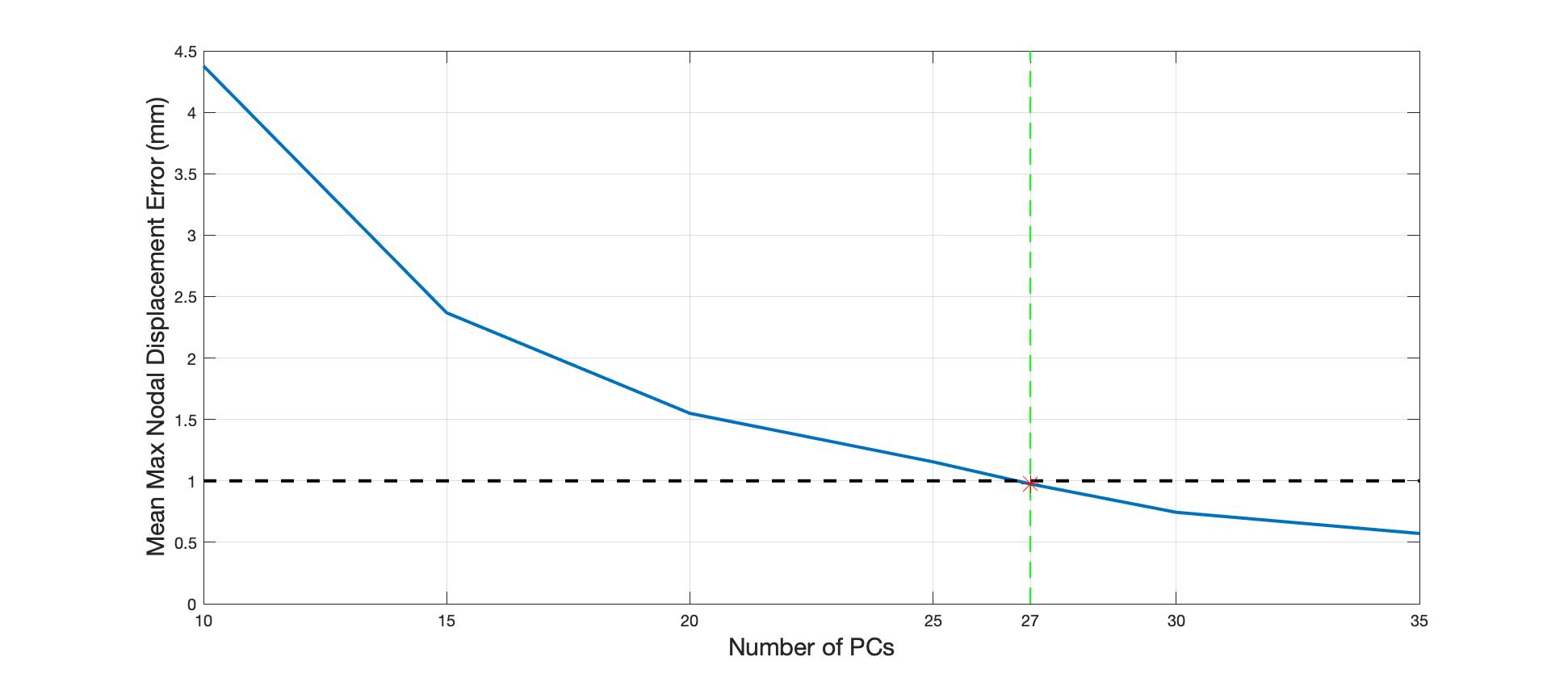}
    \caption{\label{fig:pc_opt} Maximum nodal displacement error across entire data set for varying number of principal components used during reconstruction.
    }
\end{figure}

The number of principal components $n_w$ is selected as the minimum number of principal components required for the maximum nodal displacement error to be less than $1~mm$ when using PCA reconstruction. Figure \ref{fig:pc_opt} shows this error versus the number of PCs used in reconstruction. The displacement error is the average maximum error for 5-fold cross validation across the entire data set.

The parametric study regarding $n_{d}$ is conducted on $G_{test1}$. Table \ref{tab:parameterization_vs_FM_num} shows the parameterization results with respect to different $n_{d}$s ranging from 3 to 20. The results are reported in the average of $\mathrm{Offset_{mean}}$ and the average of $\mathrm{Offset_{max}}$ of all testing samples. 

\begin{table}[!ht]
  \caption{Parameterization results w.r.t. $n_{d}$. }
  \label{tab:parameterization_vs_FM_num}
  \centering
  \begin{tabular}{ccc}
        \toprule
        $n_{d}$	& Average of $\mathrm{Offset_{max}}$ (mm) & Average of $\mathrm{Offset_{mean}}$ (mm) \\
        \midrule
        $3$ & 2.205 & 0.634 \\
        $4$ & 1.587 & 0.593 \\ 
        $5$ & 0.837 & 0.287 \\
        $6$ & 0.729 & 0.241 \\
        $7$ & 0.675 & 0.228 \\
        $8$ & 0.641 & 0.217 \\ 
        $9$ & 0.590 & 0.203 \\
        $10$ & 0.485 & 0.176 \\
        $11$ & 0.452 & 0.167 \\
        $12$ & 0.458 & 0.168 \\
        $13$ & 0.412 & 0.159 \\
        $14$ & 0.440 & 0.162 \\
        $15$ & 0.392 & 0.148 \\
        $16$ & 0.386 & 0.144 \\
        $17$ & 0.402 & 0.153 \\
        $18$ & 0.352 & 0.136 \\
        $19$ & 0.365 & 0.141 \\
        $20$ & 0.302 & 0.113 \\
        \bottomrule
  \end{tabular}
\end{table}

Results of the aforementioned parametric studies show that with a few number of FMs and PCs, the pre-trained ANN can reconstruct the deformed configuration with a very high euclidean accuracy. With less number of FMs, it can be practically more convenient to track the displacements of a few points inside a soft tissue; with less number of PCs, the ANN enables a higher quality nonlinear mapping between the FM displacements and the weight vectors, which facilitates the learning process and benefits the speed and accuracy of deformation reconstruction.


\end{document}